\documentclass{moriond}

\usepackage{slashed}
\usepackage{amsmath,amssymb}
\usepackage[compat=1.1.0]{tikz-feynman}
\usepackage{multirow}
\usepackage{lineno}
\usepackage[symbol]{footmisc}

\def\be{\begin{equation}}
\def\ee{\end{equation}}
\def\bea{\begin{eqnarray}}
\def\eea{\end{eqnarray}}

\newcommand{\Photo}{}

\begin{document}
\vspace*{4cm}
\title{CONFRONTING MINIMAL FREEZE-IN MODELS WITH THE LHC}

\author{G. B\'ELANGER$^{\, \mathrm{a}}$, N. DESAI$^{\, \mathrm{b}}$, A. GOUDELIS$^{\, \mathrm{c}}$, {\bf J. HARZ}$^{\, \mathrm{d},}$~\footnote[1]{Speaker}, A. LESSA$^{\, \mathrm{e}}$, J.M. NO$^{\, \mathrm{f}}$, A. PUKHOV$^{\, \mathrm{g}}$, S. SEKMEN$^{\, \mathrm{h}}$, D. SENGUPTA$^{\, \mathrm{i}}$, B. ZALDIVAR$^{\, \mathrm{a,f}}$, J. ZURITA$^{\, \mathrm{j,k}}$}

\address{
\vspace{4mm}
$^{\, \mathrm{a}}$LAPTh, Univ. Grenoble Alpes, USMB, CNRS,  74940 Annecy, France
\\[3pt]
$^{\, \mathrm{b}}$L2C and LUPM, CNRS-Universit\'e de Montpellier, France
\\[3pt]
$^{\, \mathrm{c}}$Sorbonne Universit\'e, CNRS, Laboratoire de Physique Th\'eorique et Hautes \'Energies, LPTHE, F-75252 Paris, France and Sorbonne
Universit\'es, Institut Lagrange de Paris (ILP), 98 bis Boulevard Arago, 75014 Paris, France
\\[3pt]
$^{\, \mathrm{d}}$Physik  Dept.  T70,  James-Franck-Str., Technische  Universit{\"a}t  M{\"u}nchen, 85748  Garching,  Germany
\\[3pt]
$^{\, \mathrm{e}}$CCNH, Universidade Federal do ABC, Santo Andr\'e, 09210-580 SP, Brazil
\\[3pt]
$^{\, \mathrm{f}}$Dpto. and Instituto de Fisica Teorica, IFT-UAM/CSIC,
Cantoblanco, 28049, Madrid, Spain
\\[3pt]
$^{\, \mathrm{g}}$Skobeltsyn Institute of Nuclear Physics, Moscow State University, Moscow 119992, Russia
\\[3pt]
$^{\, \mathrm{h}}$Kyungpook National University, Physics Dept., Daegu, South Korea
\\[3pt]
$^{\, \mathrm{i}}$Department of Physics and Astronomy, Michigan State University, East Lansing, MI, U.S.A
\\[3pt]
$^{\, \mathrm{j}}$IKP, Karlsruhe Institute of Technology, Hermann-von-Helmholtz-Platz 1, D-76344 Eggenstein-Leopoldshafen, Germany
\\[3pt]
$^{\, \mathrm{k}}$TTP, Karlsruhe Institute of Technology, Engesserstra{\ss}e 7, D-76128 Karlsruhe, Germany
}

\maketitle\abstracts{
We present a class of dark matter models, in which the dark matter particle is a feebly interacting massive particle (FIMP) produced via the decay of an electrically charged and/or colored parent particle. Given the feeble interaction, dark matter is produced via the freeze-in mechanism and the parent particle is long-lived. The latter leads to interesting collider signatures. We study current LHC constrains on our models arising from searches for heavy charged particles, disappearing tracks, displaced leptons and displaced vertices. We demonstrate not only that collider searches can be a powerful probe of the freeze-in dark matter models under consideration, but that an observation can lead as well to interesting insights on the reheating temperature and thus on the validity of certain baryogenesis models.}

\section{Introduction}
For many years already, the so called Weekly Interacting Massive Particle (WIMP) miracle has guided (astro)particle physics in its endeavor to identify the nature of dark matter (DM), naively expecting a WIMP mass around the electroweak scale. However, as neither direct or indirect detection experiments, nor the LHC has provided any hint for a WIMP so far, new directions are currently being pursued. One possible explanation for the non-observation of DM so far could be that DM is a Feebly Interacting Massive Particle (FIMP)~\cite{Hall:2009bx}. In contrast to a WIMP that is in thermal equilibrium at early times and creates the DM abundance via the freeze-out mechanism, a FIMP is assumed \textit{not} to be in equilibrium at early times. It gets produced via the decay of a parent particle $F$ that has sizable couplings to the SM. Once the temperature falls below $m_F$, the production gets suppressed and the DM abundance \textit{freezes-in}. Due to its feeble interaction to the DM particle $s$, $F$ occurs to be a long lived particle (LLP) and leads to a variety of non-prompt signatures at collider experiments~\cite{Alimena:2019zri}. Via a bottom-up approach, we defined a class of models~\cite{Belanger:2018sti} that is motivated by the least number of exotic fields for successful DM freeze-in but can, at the same time, be explored by collider signatures. We performed a recasting of the relevant LLP searches for our freeze-in models.

\section{The class of models}
We extended the SM by a real scalar DM candidate $s$, a singlet under $SU(3)_c \times SU(2)_L \times U(1)_Y$, and a vector-like fermion $F$, a singlet under $SU(2)_L$. Both $s$ and $F$ are chosen to be odd under a $Z_2$ symmetry. The corresponding Lagrangian reads
\begin{align}
\label{eq:lagrangian}
{\cal{L}}  &= {\cal{L}}_{\rm SM} + \partial_\mu s ~ \partial^\mu s  - \frac{\mu_s^2}{2} s^2 + \frac{\lambda_s}{4} s^4 + \lambda_{sh} s^2 \left(H^\dagger H\right) \\ \nonumber
 &+ \bar{F} \left(i\slashed{D}\right) F  - m_{F} \bar{F} F - 
\sum_{f} y_{s}^{f} \left(s \bar{F} \left( \frac{1+\gamma^5}{2} \right) f + {\rm{h.c.}} \right),
\end{align}
with $f = \lbrace e, \mu, \tau \rbrace$, $\lbrace u, c, t \rbrace$ or $\lbrace d, s, b \rbrace$ being the right-handed SM fermion. This leads to three possible models with the left-handed component of the vector-like fermion $F$ transforming as $(1,1,-1)$ (leptonic model) or $(3,1,-2/3)$ and $(3,1,1/3)$ (hadronic model) under $SU(3)_c \times SU(2)_L \times U(1)_Y$. In the following, we will only consider the up-type case for the hadronic model and generally neglect couplings to the third generation fermions. Moreover, we set the DM self-coupling $\lambda_{s}=0$ and the Higgs portal term $\lambda_{sh}=0$~\cite{Belanger:2018sti}. Hence, we are left with three free parameters $m_s, m_F, \{y_s^f\}$. We assume $m_s < m_{F}$ and a feeble coupling $\{y_s^f\} \sim \mathcal{O}(10^{-13}-10^{-7})$.
We implemented this class of models in {\tt FeynRules}. The corresponding model files can be downloaded from~\cite{FICPLHC}.

\section{Cosmological and indirect constraints}
{\bf Relic density.} In our set-up, DM is mainly produced via the decay of a vector-like particle $F$. We make the assumption that the initial DM density is $n_s=0$ and DM is produced during radiation domination. Thus, the DM yield $Y_s$ is given by
\begin{align}
Y_s \approx \dfrac{90 \, M_{\rm Pl}}{8\pi^4\cdot 1.66} \dfrac{g_F}{m_F^2}\,\Gamma \,\int_{m_F/T_R}^{m_F/T_0} dx~x^3 \dfrac{K_1(x)}{g^s_*(m_F/x)\sqrt{g_*(m_F/x)}},
\label{eq:DMyield}
\end{align}
with $g_F$ being the internal degrees of freedom of $F$, and $g_*$ and $g^s_*$  the effective degrees of freedom for the energy and entropy densities, respectively. The temperature today is denoted by $T_0$, $T_R$ is the reheating temperature, and  $x=m_F/T$. The function $K_1(x)$ is the modified Bessel function of the second kind. By assuming that $s$ contributes to the DM abundance, we can relate the decay length $c\tau$ to the particle masses $m_s$ and $m_F$, and the reheating temperature $T_R$,
\begin{align}
c\tau \approx 9~{\rm m}~ g_F \left(\dfrac{0.12}{\Omega_s h^2}\right)
\left(\dfrac{m_s}{100{\rm keV}}\right)\left(\dfrac{200{\rm GeV}}{m_F}\right)^2
\left(\dfrac{102}{g_*(m_F/3)}\right)^{3/2}
\left[\dfrac{\int_{m_F/T_R}^{m_F/T_0} dx~x^3 K_1(x)}{3\pi/2}\right]~.
\label{eq:ctau}
\end{align}
The relation above indicates that for $m_F \ll T_R$, long particle lifetimes are expected.\\

{\bf Lyman-alpha forest and Big Bang Nucleosynthesis.} 
However, if $F$ decays too late, it may effect the predictions from Big Bang Nucleosynthesis (BBN). In our collider analysis we focused on decay lengths in the range $[1\mathrm{cm}, 10^4\mathrm{m}]$, where the longest corresponds to a temperature of roughly $150~\mathrm{MeV}$. This means that the decay happens much before the onset of BBN and will not alter the history of BBN.

Besides, DM with a non-negligible velocity dispersion can possibly lead to the washout of small scale structures. The most stringent limit results from the Lyman-$\alpha$ forest and we found a lower bound on the DM mass of $m_s \gtrsim 12~\mathrm{keV}$ ~\cite{Belanger:2018sti,Boulebnane:2017fxw}.\\

{\bf Indirect constraints.} 
Additionally, we checked for different indirect constraints. For the leptonic case, we can exclude any violation of the muon life time~\cite{Tanabashi:2018oca} or lepton flavour violation observables~\cite{TheMEG:2016wtm}, as the DM particle generally interacts too feebly to make any significant impact. Furthermore, we do not expect any relevant contributions to electroweak precision observables, as our vector-like fermion $F$ does not mix with the SM fermions and is a $SU(2)_L$ singlet~\cite{Ellis:2014dza}. For the hadronic case, we can further exclude any violation of current limits of meson-mixing or rare meson decay constraints again due to the feeble interaction $F-f-s$. Moreover, effects on the running of the strong coupling are not expected for masses of a few hundred GeV~\cite{Llorente:2018wup}.

\section{Collider constraints}
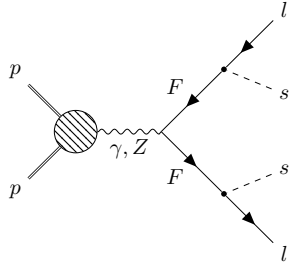
\begin{figure}
\centering
\scalebox{0.75}{
\begin{tikzpicture}
\begin{feynman}
  \node [blob] (a);
  \vertex [above left=of a] (i1) {$p$};
  \vertex [below left=of a] (i2) {$p$};
  \vertex [right=of a] (b);
  \node [above right=of b,circle,fill,inner sep=1pt] (E1);
  \node [below right=of b,circle,fill,inner sep=1pt] (E2);
  \vertex [above right=of E1] (l1) {$l$};
  \vertex [below=of l1] (s1) {$s$};
  \vertex [below right=of E2] (l2) {$l$};
  \vertex [above=of l2] (s2) {$s$};

\diagram*{
   (i1) -- [double] (a) -- [double] (i2),
   (E1) -- [fermion,edge label'={$F$}] (b) -- [fermion,edge label'={$F$}] (E2),
   (s1) -- [scalar] (E1) -- [anti fermion] (l1), 
   (s2) -- [scalar] (E2) -- [fermion] (l2),
   (a) -- [photon,edge label'={$\gamma,Z$}] (b),
};
\end{feynman}
\end{tikzpicture}}
\caption{\it Drell-Yan pair-production of $F$ and its subsequent decay.}
\label{fig:LHC_production}
\end{figure}

For the leptonic case, as depicted in Fig.~\ref{fig:LHC_production}, the collider signature proceeds via Drell-Yan pair-production of $F$ and a subsequent decay into two leptons and two scalar singlets. Prompt searches at LEP2 put a bound on $m_F > 104~\mathrm{GeV}$ for the leptonic case. It has been shown that prompt searches do not compete with dedicated LLP searches for $c\tau \gtrsim 0.5~\mathrm{cm}$. For the hadronic process, the production proceeds via $s$-channel gluon exchange or $t$-channel exchange of $F$ fermions and we similarly do not expect to be constrained by prompt searches.

Due to the feeble $F-f-s$ coupling, the $F \rightarrow f s$ decay can be displaced or even happen outside the detector. Depending on the lifetime and thus on the decay length of our LLP $F$, we can distinguish between three main search strategies for both the leptonic and hadronic model:\\

{\bf Heavy Stable Charged Particles (HSCP).}
If the charged particle $F$ is sufficiently long lived, it will decay after having passed as \textit{Heavy Stable Charged Particle (HSCP)} through the tracker (or even the muon chambers). Hereby, the HSCP signature depends strongly on the specific nature of $F$. As colour-neutral particle in the leptonic model, it would appear as ionizing track, while as colour-charged one in the hadronic model, it would hadronize into neutral and/or charged $R$-hadrons. Being heavier than a SM particle, a HSCP can be distinguished from background by its higher ionziation energy loss and, when decaying outside the detector, by a longer time-of-flight (TOF) in the muon chambers in contrast to relativistic muons. Thus, we distinguish between a \textit{tracker-only analysis} with decays happening after the tracker and a \textit{tracker+TOF analysis} with the HSCP decaying after the muon chamber. The expected signatures of our models are comparable to those in~\cite{Chatrchyan:2013oca,CMS:2016ybj}, in which limits on staus and stops were set in a gauge mediated SUSY breaking model. We have recasted the limits of these two CMS analyses ($8~\mathrm{TeV}$ analysis with $18.8~\mathrm{fb}^{-1}$~\cite{Chatrchyan:2013oca} and $13~\mathrm{TeV}$ analysis with $12.9~\mathrm{fb}^{-1}$~\cite{CMS:2016ybj}).  To account for a finite lifetime of $F$, we rescaled our production cross section of $F$ (which was computed with {\sc MadGraph5\_aMC@NLO}). For further details, we refer to~\cite{Belanger:2018sti}.
For the leptonic model, we found that the $13~\mathrm{TeV}$ tracker+TOF data is more constraining for larger $c\tau \approx \mathcal{O}(10\mathrm{m}-100\mathrm{m})$, while the  $8~\mathrm{TeV}$ tracker-only data is more constraining for smaller $c\tau \approx \mathcal{O}(1\mathrm{m})$ due to more integrated luminosity. As $R$-hadrons can flip their charge when traversing the detector, the tracker-only selection may fail the tracker+TOF selection. Thus, the tracker-only analysis leads to the most stringent constraints for the hadronic model.\\

{\bf Disappearing tracks (DT).} 
For medium life times, searches for \textit{disapearing tracks (DT)} are relevant. Hereby, the LLP decays within the innermost tracker, while other decay products either escape the detector without interaction (DM candidate) or are too soft to be reconstructed. For our purposes we recasted two analyses of $13~\mathrm{TeV}$ data of ATLAS ($36.1~\mathrm{fb}^{-1}$)~\cite{Aaboud:2017mpt} and CMS ($138.4~\mathrm{fb}^{-1}$)~\cite{Sirunyan:2018ldc} for an AMSB motivated scenario with mass degenerate lightest charginos and neutralinos. The object to be identified is an isolated track reconstructed in the pixel and strip detectors without any hit in the outer tracker (CMS) or a track with only hits in the pixel detector (ATLAS).
For further details on the selection criteria, we refer to~\cite{Belanger:2018sti}. However, we want to note that ATLAS requires no activity in the muon system, such that our ATLAS limits correspond to $\mathrm{BR}(F\rightarrow es) =1.0$.
For recasting, generator-level efficiency tables depending on the resonant mass and the lifetime are provided by the experimental collaborations ~\cite{Aaboud:2017mpt,Sirunyan:2018ldc}. The number of events passing the selection criteria are obtained by the product of the production cross section times efficiency and luminosity ($\mathcal{N} = \sigma_\mathrm{pp\rightarrow F \bar F} \times \varepsilon(m,\tau) \times \mathcal{L}$). As expected by the general experimental features, our results show that ATLAS leads to stronger exclusion limits for smaller decay lengths (e.g. excluding $m_F < 275~\mathrm{GeV}$ for $c\tau \sim 20 \mathrm{cm}$), while CMS constrains more efficiently larger decay lengths (e.g. excluding $m_F < 335~\mathrm{GeV}$ for $c\tau \sim 1 \mathrm{m}$).\\

{\bf Displaced lepton searches (DL).}
In case of the leptonic model and smaller expected lifetimes, the decay $F\rightarrow \ell s$ is expected to take place as \textit{displaced lepton (DL)} decay within the detector.
In that case, the CMS searches for non-prompt $R$-parity violating SUSY decays (e.g. $\tilde{t}_1 \rightarrow b \ell$) with oppositely charged, displaced muons and electrons at $8~\mathrm{TeV}$ with $19.7~\mathrm{fb}^{-1}$~\cite{Khachatryan:2014mea} and at $13~\mathrm{TeV}$ with $2.6~\mathrm{fb}^{-1}$~\cite{CMS:2016isf} can constrain our model. As these searches are maximally sensitive for $y_s^e \approx y_s^\mu$, our results depend on the final branching ratios.
For the event selection exactly one muon and one electron is required.
The decay of the LLP has to happen within a distance of $L_z < 300$ mm and $\sqrt{L_x^2 + L_y^2} < 40$ mm. Further details on the selection criteria are given in~\cite{Belanger:2018sti}.
We used {\sc MadGraph5\_aMC@NLO} to simulate the production of $F$ and its subsequent decay considering different branching ratios for the decay into muons or electrons. We generated 200k MC events for each set of $\{m_F, c\tau\}$, applied the provided efficiencies and the CMS event selection, in order to obtain the 95 \% C.L. exclusion limits for our leptonic model.\\

{\bf Displaced vertices plus MET (DV+MET).}
In case of the hadronic model and smaller expected lifetimes, \textit{displaced jets with missing energy} can be searched for. ATLAS performed an analysis with 13~TeV data and 32.8~fb$^{-1}$ for a simplified split SUSY model, in which a long-lived gluino hadronizes into an $R$-hadron and decays subsequently ~\cite{Aaboud:2017iio}. As selection criteria, multi-track ($\geq 5$) displaced vertices in association with large missing transverse momenta ($E_T^{miss} > 250$~GeV) and a visible invariant mass greater than 10 GeV were required. We used  {\tt MadGraph 5} and Pythia 8 to simulate the production and decay of $F$. For hadronization, we used the Pythia 8 hadronization model for long-lived stops. We generated 50k MC events for each set of $\{m_F, c\tau\}$. In order to account for detector response and event reconstruction, we applied the efficiencies that are given as a function of vertex radial distance, number of tracks and mass ~\cite{Aaboud:2017iio} to our truth signal MC events.

\section{Link to baryogenesis}
In order to generate the baryon asymmetry of our universe (BAU), the three Sakharov conditions have to be fulfilled: baryon number ($B$) violation, $C$ and $CP$ violation have to occur out of equilibrium. Only $B$ violation is sufficiently realised in the SM via sphaleron transitions which are only active above a certain critical temperature $T^*$. Many models that explain the BAU, e.g. leptogenesis, rely on active sphaleron interactions. In leptogenesis a lepton asymmetry is firstly generated and then translated into a BAU via sphalerons. However, only if the lepton asymmetry is generated above the critical temperature $T^*$, sphalerons are active and can generate the BAU. Comparing with Eq.~\ref{eq:ctau}, we see that in case of an observation of a particle $F$ with mass $m_F$ and decay length $c \tau$, we are left with the two free parameters $m_s$ and $T_R$. Under the assumption that $m_s$ contributes to DM, we can constrain the reheating temperature $T_R$. For a conservative estimate, we assume the lightest possible mass according to constraints from the Lyman-$\alpha$ forest, $m_s=12~\mathrm{keV}$, and can hence identify $T_R$ under the assumption that $s$ makes up the full DM density. Note, that a higher DM mass $m_s > 12~\mathrm{keV}$ or the possibility that $s$ only makes up a part of the DM abundance, will imply an even smaller $T_R$. If a possible observation indicates that $T_R < T^*$, sphaleron interactions will not be active. This would allow to exclude baryogenesis and leptogenesis models with a critial temperature $T^*$ that rely on active sphaleron transitions.

\section{Conclusions}
\begin{figure}[t]
\centering
\includegraphics[width=0.8\textwidth]{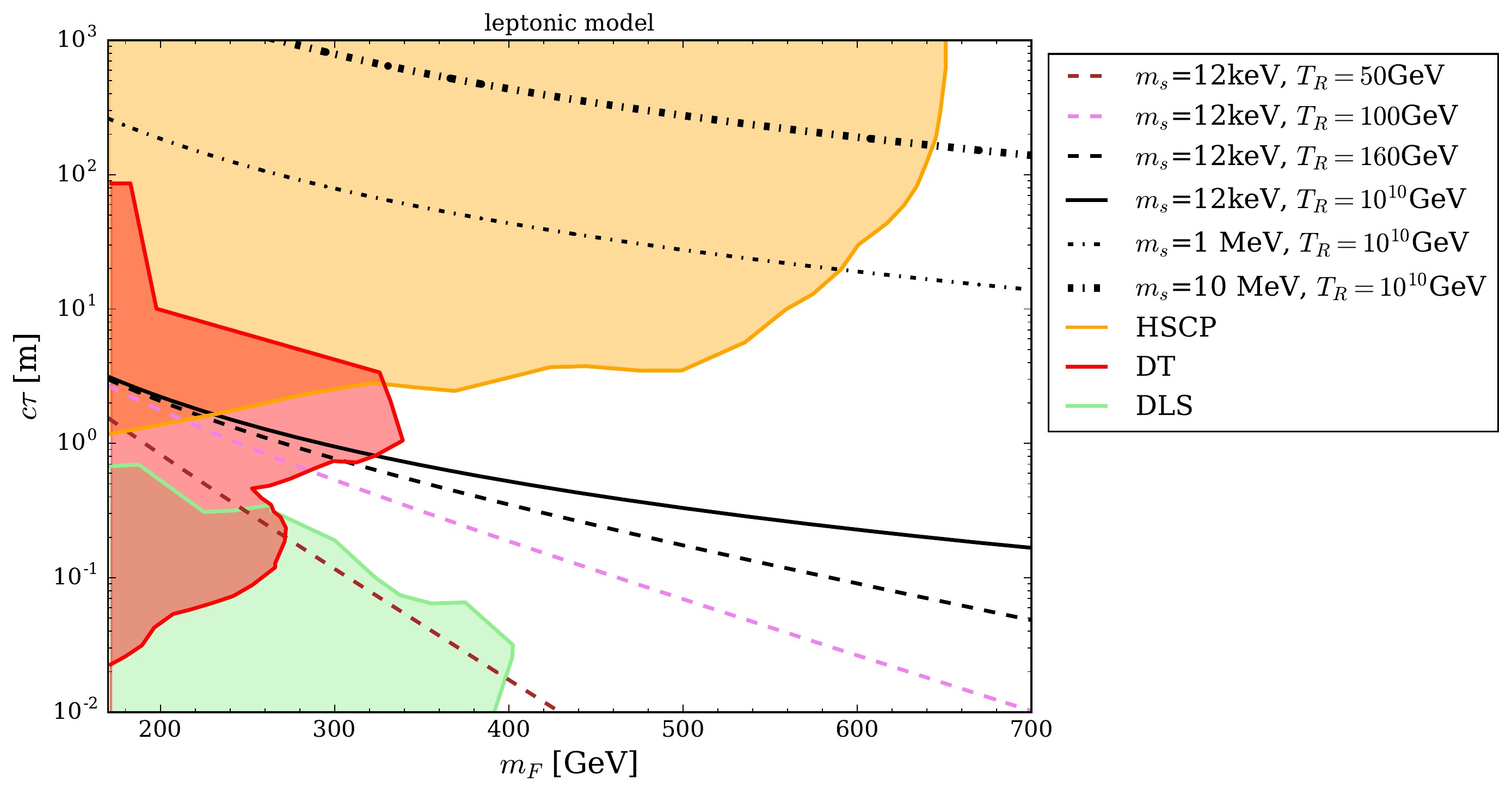}
\caption{Summary of the LHC constraints for the lepton-like FIMP scenario.}
\label{fig:leptonic}
\end{figure}

\begin{figure}[t]
\centering
\includegraphics[width=0.8\textwidth]{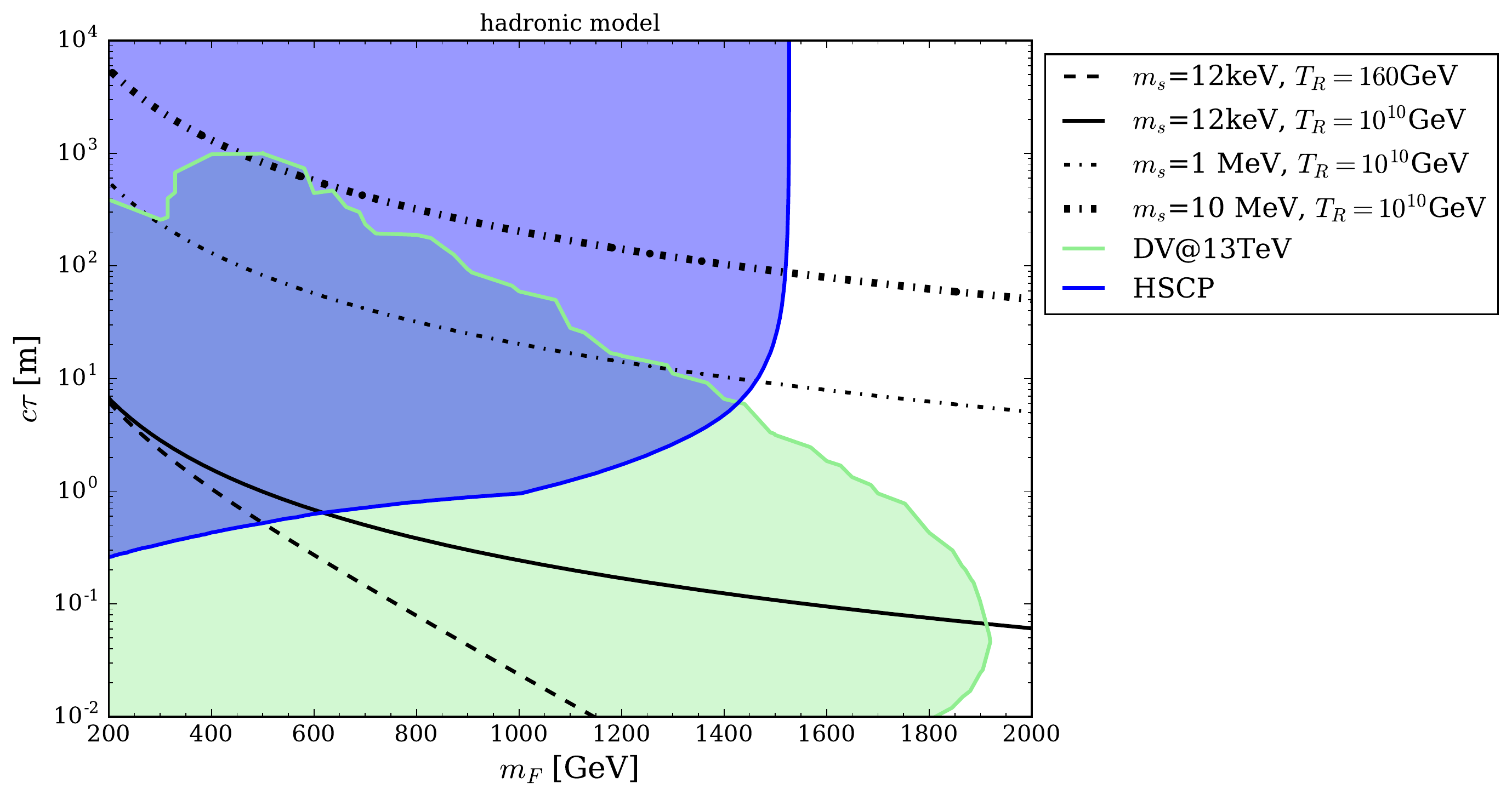}
\caption{Summary of the LHC constraints for the quark-like FIMP scenario.}
\label{fig:hadronic}
\end{figure}

In Fig.~\ref{fig:leptonic} and Fig.~\ref{fig:hadronic}, we summarize our results for the leptonic and hadronic model, respectively. For each type of analysis, we have created a single $95~\%$ C.L ``envelope'' exclusion line, considering the largest exclusion interval for $c\tau$ for each $m_F$. For the detailed results of the single analyses we refer to ~\cite{Belanger:2018sti}.
For comparison with possible model parameter values, we fixed the reheating temperature to $T_R=10^{10}$ GeV, much higher than all other scales in our model. For the mass of the DM particle $s$, we have chosen three values $m_s=\{12~\mathrm{keV}, 1~\mathrm{MeV}, 10~\mathrm{MeV}\}$. The black curves in Figs.~\ref{fig:leptonic} and \ref{fig:hadronic} show the corresponding lines in the parameter space $\{m_F,c\tau\}$ for which the observed DM abundance is matched. When fixing the DM mass to the lightest possible value $m_s=12~\mathrm{keV}$, we show for three additional reheating temperatures $T_R=\{50, 100, 160\}$ GeV the corresponding line in parameter space for the leptonic model. This demonstrates that when measuring a particle $F$ with mass $m_F$ decaying with a decay length $c\tau$, we can extract the highest reheating temperature possible. Note, that every DM mass heavier than $m_s=12~\mathrm{keV}$, would only lead to an even lower reheating temperature. If we find $T_R<T^*$, with $T^*$ being the critical temperature at which sphalerons cease to be active in a specific model, we can falsify these baryogenesis models.  For the SM, $T^* = (131.7 \pm 2.4)~\mathrm{GeV}$~\cite{DOnofrio:2014rug}, while other models can have different critical temperatures, e.g.  a supercooled scenario can feature a critical temperature as low as $T^* \sim 50~\mathrm{GeV}$. As this interesting area is already excluded by displaced vertex searches for the hadronic model, we show only $T_R=160~\mathrm{GeV}$ in Fig.~\ref{fig:hadronic}. Similarly, we have performed an extrapolation for the high luminosity LHC (HL-LHC), for which we refer to~\cite{Belanger:2018sti}. We found that the HL-LHC will be able to almost fully cover the parameter space of the leptonic model for $T_R \leq 160~\mathrm{GeV}$ by displaced lepton searches, except of a small area for $m_F \gtrsim 800~\mathrm{GeV}$ and very small decay lengths.

This study clearly demonstrates the impact of LLP searches at the LHC for freeze-in models. Especially, with the absence of DM signals from direct and indirect detection experiments, this avenue of research is promising to explore in the near future. 

\section*{Acknowledgments}
The speaker wants to thank her collaborators for the great collaboration and the organizers of Moriond EW 2019. JH  was supported by the DFG Emmy Noether Grant No. HA 8555/1-1. For all further funding acknowledgements, we refer to the original publication~\cite{Belanger:2018sti}.

%

\end{document}